\documentclass{hep99}

\def\beq{\begin{equation}}
\def\eeq{\end{equation}}
\def\beqa{\begin{eqnarray}}
\def\eeqa{\end{eqnarray}}
\newcommand{\eq}[1]{Eq.~(\ref{#1})}
\newcommand{\nl}{\nonumber \\}

\begin{document}

\title{Evolution of parton distributions with truncated Mellin moments}

\author{Lorenzo Magnea$^1$ and Stefano Forte$^2$}

\address{$^1$Dipartimento di Fisica Teorica, Universit\`a di Torino, 
and INFN, Sezione di Torino, \\ Via P. Giuria 1, I-10125, Torino, ITALY\\
$^2$INFN, Sezione di Roma III, Via della Vasca Navale 84, I-00146, Roma, 
Italy \\ (on leave from INFN, sezione di Torino)\\[3pt] 
E-mail: {\tt magnea@to.infn.it}}

\abstract{Evolution equations for parton distributions can be approximately
diagonalized and solved in moment space without assuming any knowledge of the 
parton distribution in the region of small $x$. The evolution algorithm for
truncated moments is simple and rapidly converging. 
Examples of applications are outlined.}

\maketitle


\section{Introduction \label{in}}

Altarelli--Parisi evolution equations for parton distributions can be 
exactly diagonalized and analytically solved by taking a Mellin transform,
which turns the $x$-space integro-differential equation into a set of 
decoupled ordinary first order differential equations. This procedure, 
although elegant, suffers from a practical drawback: Mellin moments are 
integrals over the entire allowed kinematical range for the parton momentum 
fraction, $0 < x < 1$, whereas in practice parton distributions are measured
only in a finite subinterval of that range, say $x_0 < x < x_1$. Furthermore,
solving the equation through a Mellin transform somewhat obscures the 
directionality in $x$-space of parton evolution, namely the fact that the 
evolution in $Q^2$ of the parton distribution at a given value of the 
momentum fraction $z$ depends only on the values of the distribution for 
$x > z$. In this note I present a method \cite{us} to evolve parton 
distributions in Mellin space without using any assumptions on the 
small--$x$ behavior of parton distributions.
The method is based on the notion of ``truncated'' moments, defined as 
integrals over a region $x_0 < x < 1$. It turns out that (non--singlet)
truncated moments are coupled by evolution via a triangular matrix whose 
effective size can be kept small without loss of accuracy, and that can be 
diagonalized analitycally with little effort; numerical implementation is thus 
relatively simple and efficient. The method addresses a specific 
source of uncertainty in the determination of parton distributions (the lack
of small--$x$ data, which are abundant only for $F_{2,p}(x,Q^2)$). This 
uncertainty is apparent in Mellin space, but is present also in 
$x$--space evolution codes, which operate by choosing a fixed parametrization 
for the parton distribution and subsequently fitting the parameters to the 
data. Any parametrization has a degree of rigidity and theoretical prejudice, 
so that tacit assumption made about the $x \to 0$ region affect also the region
where the data are fitted. This kind of uncertainty, in principle very 
difficult to quantify, is replaced in the present method by a small, 
quantifiable and sistematically reducible uncertainty due to the finite 
size chosen for the triangular matrix of couplings between truncated moments.

\section{Truncated moments \label{tm}}

For the sake of simplicity, in the following I will consider the nonsinglet
quark distribution, $q(x,Q^2)$. The extension of the method to singlet and 
gluon distributions requires some work, which has recently been 
completed \cite{us2}. Consider the evolution equation
\beq
\frac{d}{dt}~q(x) =
\frac{\alpha_s}{2 \pi} \int_x^1 \frac{d y}{y} 
P\left(\frac{x}{y}\right) q(y)~~,
\label{alpar}
\eeq
where the evolution kernel $P(z)$ is a perturbative series in $\alpha_s(Q^2)$,
and all dependence on $Q^2$ has been suppressed.
Define the truncated moment of $q(x)$ as
\beq
q_N(x_0) \equiv \int_{x_0}^1 d x x^{N - 1} q(x)~~.
\label{trunc}
\eeq
Then the evolution equation for truncated moments reads
\beq
\frac{d}{dt} q_N(x_0) =
\frac{\alpha_s}{2 \pi} \int_{x_0}^1 d y y^{N - 1} q(y) 
G_N\left(\frac{x_0}{y}\right)~~,
\label{truncalpar}
\eeq
where
\beq
G_N(x) = \int_x^1 d z z^{N - 1} P(z)~~.
\label{kern}
\eeq
It is easy to show that the $N$-th truncated moments is coupled by 
\eq{truncalpar} to all moments with index $N + K$, $(K >0)$. Furthermore, 
the strength of this coupling decreases rapidly with $K$. To prove it, one
expands $G_N(x_0/y)$ in Taylor series around $y = 1$. This expansion has 
radius of convergence $1 - x_0$, and can be truncated, say at order $M$.
Reorganizing the polynomial in $y - 1$ thus obtained to collect the 
different powers of $y$, and performing the $y$ integration, one ends up with
an equation coupling the $N$-th moment to the subsequent $M$ moments,
\beq
\frac{d}{dt} q_N(x_0) =
\frac{\alpha_s}{2 \pi} \sum_{K=0}^M c_{K,N}^{(M)}(x_0) 
q_{N + K}(x_0)~~,
\label{finsyst}
\eeq
with coefficients $c_{K,N}^{(M)}(x_0)$ that are perturbatively calculable
for arbitrary $x_0$, given the evolution kernel $P(z)$. For moderately small 
$x_0$, say $x_0 \leq 0.1$, one gets a very good approximation to the exact 
evolution for reasonably small values of $M$, say $M \sim 3 - 5$, and 
furthermore the value of $M$ required to keep the desired accuracy decreases 
with $N$. This can be understood by noting that, for $x_0 = 0$, $G_N(x_0/y)$ is
independent of $y$, so that for finite but small $x_0$, and over most of the 
$y$ integration range, one is Taylor expanding a function which is 
approximately a constant. Furthermore, the contribution to the evolution 
of higher moments is suppressed by the fact that they are dominated by 
high values of $x$, where the parton distribution themselves are decreasing 
as powers of $(1 - x)$. Notice that the accuracy of these approximations for 
each truncated moment can always be controlled by comparing numerically to 
the exact evolution, given by \eq{truncalpar}. 

\section{NLO evolution \label{ev}}

Given the approximations discussed above, the evolution of a given 
truncated moment (say the $N_0$-th) can be calculated by solving a linear 
system of differential equations coupling it to moments $N_0 + 1$ through
$N_0 + M$. Schematically, at NLO,
\beq
\frac{d q_K}{d t} = \frac{\alpha_s}{2 \pi}
\sum_{L = N_0}^{N_0 + M} \left[ C^{(0)}_{K L} +
\frac{\alpha_s}{2 \pi} C^{(1)}_{K L} \right] q_L~~,
\label{simpNLO}
\eeq
a situation which is formally identical to the coupled evolution of the
gluon and singlet quark distributions. Here analytical and numerical
computations are greatly simplified by the fact that the ``anomalous 
dimension'' matrix is triangular, so that its eigenvalues are simply given 
by the diagonal entries, while the eigenvectors are given by a recursion 
relation, without the need to compute any determinants.
Denoting by $R$ the matrix diagonalizing the LO anomalous dimension $C^{(0)}$,
one can use it to rotate the set of moments $q_K$ under consideration, as 
well as the NLO matrix $C^{(1)}$, according to
\beqa
\hat{q}_K & = & \sum_{L = N_0}^{N_0 + M} R_{K L} q_L~~, \label{rot0} \\
\hat{D}_{K L} & = & \sum_{P,Q = N_0}^{N_0 + M} R_{K P}
C^{(1)}_{P Q} R^{-1}_{Q L}~~. \nonumber
\eeqa
Then the NLO solution is given by
\beqa
\hat{q}_K & = & \left[\frac{\alpha_s^0}{\alpha_s}
        \right]^{\gamma_K} \left[ 1 - \frac{\gamma_K b_1}{2 \pi b_0} \left(
        \alpha_s^0 - \alpha_s \right) \right] 
        \hat{q}_K^0\nl
&  - &  \sum_{L = N_0}^{N_0 + M} \frac{\hat{D}_{K L}}{2 \pi b_0}
        \frac{1}{\gamma_K - \gamma_L + 1} \left[
        \left(\frac{\alpha_s^0}{\alpha_s}\right)^{\gamma_L} 
        \alpha_s \right. \nl
& &     \left. - \left(\frac{\alpha_s^0}{\alpha_s}
        \right)^{\gamma_K} 
        \alpha_s^0 \right] \hat{q}_L^0~~,
\label{master}
\eeqa
where the apex $0$ denotes quantities evaluated at the input scale $Q_0$, and 
$\gamma_M = C_{M M}/b_0$.

\section{Perspectives \label{ps}}

The method presented in this note has been extended to parton distributions
with coupled evolution \cite{us2}, and can be straightforwardly extended to 
polarized partons as well. It can be applied to derive estimates for all 
physical quantities depending on evolution, unbiased by the 
extrapolation of data to small $x$. An example already studied in \cite{us}
is the running coupling itself. Other significant example include the first 
moment of the polarized structure function $g_1$, appearing in the Bjorken 
sum rule \cite{pol}, and the behavior of the unpolarized gluon distribution
at moderate to large values of $x$. Work on these applications is in 
progress.

\end{document}